\documentclass[12pt]{imsart}

\usepackage{enumitem}
\usepackage[utf8]{inputenc}
\usepackage[english]{babel}
\usepackage[margin=1in]{geometry}
\usepackage[colorlinks,citecolor=blue,urlcolor=blue]{hyperref}

\usepackage{bm}
\usepackage{graphicx}
\usepackage{amssymb,amsmath,amsthm,mathrsfs}
\usepackage{epstopdf}
\usepackage{algorithmic}
\usepackage{xcolor}
\usepackage{subfigure}
\usepackage{multirow}
\usepackage{titlesec,textcase}
\usepackage{longtable}
\usepackage{rotating}
\usepackage{booktabs}
\usepackage{siunitx} 
\usepackage{float}

\newtheorem{Theorem}{Theorem}[section]

\newtheorem{Proposition}[Theorem]{Proposition}

\newtheorem{Definition}[Theorem]{Definition}

\theoremstyle{definition}

\newcommand{\diag}{\mathrm{diag}}

\RequirePackage[normalem]{ulem} 

\definecolor{rp}{RGB}{83,54,106}

\def\boxit#1{\vbox{\hrule\hbox{\vrule\kern6pt\vbox{\kern6pt#1\kern6pt}\kern6pt\vrule}\hrule}}

\begin{document}
\begin{frontmatter}
\title{Hypothesis testing  for the uniformity of  random geometric graph}

\runtitle{Test uniformity of RGG }
\runauthor{ M. Yuan }
 
\begin{aug}
\author[B]{\fnms{Mingao} \snm{Yuan}\thanks{Mingao Yuan is grateful for the generous support provided by the Startup Fund of The University of Texas at El Paso.}\ead[label=e2]{myuan2@utep.edu}}

\address[B]{Department of Mathematical Sciences,
The University of Texas at El Paso, El Paso, TX, USA\\
\printead{e2}}

\end{aug}
 
\begin{abstract}
Random geometric graphs are widely used in modeling geometry and dependence structure in networks. In a random geometric graph, nodes are independently generated from some probability distribution $F$ over a metric space, and edges link nodes if their distance is less than some threshold. Most studies assume the distribution $F$ to be uniform. However, recent research shows that some real-world networks may be better modeled by nonuniform distribution $F$. Moreover, graphs with nonuniform $F$ have notably different properties from graphs with uniform $F$. A fundamental question is: given a network from a random geometric graph, is the distribution $F$ uniform or not? In this paper, we approach this question through hypothesis testing. This problem is particularly challenging due to the inherent dependencies among edges in random geometric graphs, a property not present in classic random graphs. We propose the first statistical test.   Under the null hypothesis, the test statistic converges in distribution to the standard normal distribution.  The asymptotic distribution is derived using the asymptotic theory of degenerate U-statistics with a kernel function dependent on the number of nodes. This technique is different from existing methods in network hypothesis testing problems. In addition, we present a method for efficiently calculating the test statistic directly from the adjacency matrix. We also analytically characterize the power of the proposed test. The simulation study shows that the proposed uniformity test has high power. Real data applications are also provided.
\end{abstract}



\end{frontmatter}

\section{Introduction}

A network, also known as a graph in mathematical terms, consists of interconnected objects. These objects are called nodes or vertices, and the connections between them are referred to as edges or links. 
Networks are often represented as adjacency matrices. An adjacency matrix uses 1 to indicate an edge and 0 to indicate no edge between nodes. Networks are extensively employed across a spectrum of domains, acting as a vital tool for investigating the complex structures and interactions within various systems \cite{N03,BM08,SBL13,CC12,CBL16,CGL16,ARS01,CV1,LLR15}. For example, networks are used to  optimize the performance of 5G wireless networks \cite{ARS01}; Gene-gene interactions can be elucidated through the analysis of biological networks \cite{CV1}.

Graphs represent data in high dimensions. To gain insights from graphs, it is necessary to make assumptions about their internal organization and relationships. A wide variety of random graph models have been created for this purpose \cite{DC23,SHL11,RPF13}. The most widely used and simplest random graph model is the Erd\H{o}s-R\'{e}nyi model, where edges are formed independently with a fixed probability. The $\beta$-model is used to capture and represent the information encoded in the degree sequence of a network \cite{RPF13}. 
Real-world networks frequently show the presence of hidden geometric spaces and dependencies between edges \cite{DC23,HRP08,GMPS23}. Such networks are modeled using Random Geometric Graphs (RGGs), which are variations of the classic Erd\H{o}s-R\'{e}nyi random graph \cite{DC23,HRP08,GMPS23}. Nodes in RGGs are distributed according to a distribution $F$ in a metric space, and edges connect nodes within a specified distance \cite{DC23}. Edges depend on the distance between nodes and are correlated as a result of the random latent positions of the nodes. RGGs, with their inherent geometry, generate networks that feature rich dependence and other key characteristics observed in real-world systems \cite{DC23,HRP08,KRHP09,ZLKG14}.  For example,  RGGs effectively capture key topological properties of protein-protein interaction  networks \cite{HRP08,KRHP09}; RGGs provide a model for understanding how opinions spread across a spatial network \cite{ZLKG14}.

A common assumption in RGG literature is that the nodes are distributed uniformly \cite{HRP08,KRHP09,GRK05,Z15,SSH18,GMPS23,BC23}. In other words, $F$ is assumed to be the uniform distribution.  For example, \cite{GRK05} studied the existence of a sharp threshold for all monotone properties of RGG with uniform $F$; \cite{Z15} investigated the threshold of the absence of isolated node in uniform RGGs; \cite{HRP08,KRHP09}  fitted RGGs with uniform $F$ to protein-protein interaction networks.

However, RGGs with nonuniform $F$ have recently attracted much attention \cite{HM09,HM12,G21,MMFE22,BJ23,PBGKL23}.  For instance, \cite{HM09,HM12} examined the probability of connectivity of RGGs with nonuniform $F$. The research in \cite{PBGKL23} indicated that RGGs with nonuniform distribution $F$ more accurately reflects the complexities and features observed in some real-world networks.

Moreover, RGGs with nonuniform $F$ have notably different properties from RGGs with uniform $F$. For example, the total degree of RGGs with uniform $F$ is a degenerate U-statistic, while the total degree of RGGs with nonuniform $F$ is a non-degenerate U-statistic. The asymptotic distribution of the total degree is distinct in RGGs with uniform $F$ versus those with nonuniform $F$. Therefore, statistical inference procedures for RGGs must account for whether $F$ is uniform or nonuniform, and  required theoretical techniques may differ markedly.

A natural question arises: whether, given a real-world network from RGGs, the nodes are uniformly distributed or not. In this paper, we address this problem by hypothesis testing. 
Under the null hypothesis, the distribution of nodes is uniform. Under the alternative hypothesis, the distribution is nonuniform. This hypothesis testing problem is particularly challenging, as the edges in RGGs are not independent like in classic random graphs \cite{GL17,JKL18,L16,RPF13}.
We propose the first statistical test for the hypotheses. Under the null hypothesis, the test statistic is approximately standard normal when the number of nodes is large.
The asymptotic distribution is obtained by applying asymptotic theory of degenerate U-statistics, where the kernel function depends on the number of nodes. This proof strategy is different from prior methods in network hypothesis testing problems. Moreover, the test statistic can be easily calculated based on the adjacency matrix. We also study the power of the proposed test theoretically. The simulation results demonstrate that the proposed uniformity test exhibits high power.  We also illustrate our theory through real-world data applications, showcasing its practical applicability.

This paper is organized as follows: Section 2 formally presents the random geometric graph and the hypotheses. Section 3 details the proposed uniformity test and provides the relevant theoretical results. Section 4 includes simulation studies and real-data applications. The proofs are deferred to Section 5.

\vskip 5mm

\noindent
{\bf Notation:} We adopt the  Bachmann–Landau notation throughout this paper. Given two sequences  $a_n$  and $b_n$, denote $a_n=\Theta(b_n)$ if $c_1\leq \left|\frac{a_n}{b_n}\right |\leq c_2$ for some positive constants $c_1,c_2$ and large $n$. Denote  $a_n=\omega(b_n)$ if $\lim_{n\rightarrow\infty}\left|\frac{a_n}{b_n}\right |=\infty$. Denote $a_n=O(b_n)$ if $\left|\frac{a_n}{b_n}\right |\leq c$ for some positive constant $c$ and large $n$. Denote $a_n=o(b_n)$ if $\lim_{n\rightarrow\infty}\left|\frac{a_n}{b_n}\right |=0$. Let $N(0,1)$ be the standard normal distribution. For a sequence of random variables $X_n$, $X_n\Rightarrow N(0,1)$ represents $X_n$ converges in distribution to $N(0,1)$ as $n$ goes to infinity. $\mathbb{E}[X_n]$ and $Var(X_n)$ denote the expectation and variance of a random variable $X_n$ respectively. $\mathbb{P}(E)$ denote the probability of an event $E$. Given a finite set $E$, $|E|$ represents the number of elements in $E$. Given positive integer $t$, $\sum_{i_1\neq i_2\neq\dots\neq i_t}$ means summation over all integers $i_1,i_2,\dots,i_t$ in $[n]=\{1,2,\dots,n\}$ such that $|\{i_1,i_2,\dots,i_t\}|=t$. $\sum_{i_1< i_2<\dots< i_t}$ means summation over all integers $i_1,i_2,\dots,i_t$ in $[n]$ such that $i_1<i_2<\dots<i_t$.

\section{Model and hypotheses}

A graph is a mathematical structure that models the relationship between a set of nodes. Given an integer $n$, let  $\mathcal{V}=\{1,2,\dots,n\}$.  A graph is an ordered pair $\mathcal{G}=(\mathcal{V},\mathcal{E})$, where $\mathcal{V}$ is a set of nodes and $\mathcal{E}$ is a set of a set of edges. Each edge connects two nodes in $\mathcal{V}$.  A graph can be conveniently represented as an adjacency matrix $A$. If $\{i,j\}$ is an edge, then $A_{ij}=1$. Otherwise $A_{ii}=0$. Since $\mathcal{G}$ is undirected, the adjacency matrix $A$ is symmetric.  The number of edges connecting a node is called the degree of the node.  A path in a graph is a sequence of edges which joins a sequence of vertices. For instance, the two edges $\{1,2\}$ and $\{2,3\}$ form a path of length 2, and the three edges $\{1,2\}$,  $\{2,3\}$ and $\{3,4\}$ constitute a path of length 3.

Random graph refers to probability distributions over graphs. Random graphs have been widely used to model real-world networks \cite{HRP08,KRHP09,SHL11}. The most commonly studied is the Erd\"{o}s-R\'{e}nyi random graph, where every possible edge appears independently with probability $p$. Real-world networks frequently exhibit both geometrical organization and underlying relational dependencies. The random geometric model is a popular model for this kind of networks.

\begin{Definition}\label{def1}
Let $r_n\in(0,0.5)$ be a real number, $m$ be a positive integer and $F$ be a probability distribution on the unit square $[0,1]^m$. Given i.i.d. random variables $X_1,X_2,\dots,X_n\sim F$, the Random Geometric Graph $\mathcal{G}_{n,m}(r_n,F)$ is defined as follows:
\[A_{ij}=I[d(X_i,X_j)\leq r_n],\]
 where $A_{ii}=0$, $X_i=(X_{i1},X_{i2},\dots,X_{im})$ and 
 \begin{eqnarray}\label{distance}
 d(X_i,X_j)=\max_{1\leq k\leq m}\Big\{\min\{|X_{ik}-X_{jk}|,1-|X_{ik}-X_{jk}|\}\Big\}.
 \end{eqnarray}
\end{Definition}

The unit square $[0,1]^m$ equipped with the distance (\ref{distance}) can be described as the $m$-dimensional torus. The distance (\ref{distance}) is called the $\infty$-norm and $L_{\infty}$ distance on the torus in  \cite{BKL19} and \cite{BB24} respectively. When $F$ is uniform, $\mathcal{G}_{n,m}(r_n,F)$ is the $L_{\infty}$ random geometric graph studied in \cite{BB24}. In $\mathcal{G}_{n,m}(r_n,F)$, each vertex is  independently sampled from the distribution $F$ on $[0,1]^m$ and two nodes are connected by an edge if their distance is less than $r_n$.
 The parameter $r_n$ therefore models the sparsity of the graph. A graph with larger $r_n$ has more edges on average. When $r_n\geq 0.5$, the graph is complete. The edges in $\mathcal{G}_{n,m}(r_n,F)$ are correlated and depend on the distance between nodes. This makes the random geometric graph $\mathcal{G}_{n,m}(r_n,F)$ suitable for modeling geometry and dependence structure in real-world networks \cite{DC23,GMPS23,HRP08,KRHP09}. 

We call $\mathcal{G}_{n,m}(r_n,F)$ non-uniform (or uniform) random geometric graph if $F$ is non-uniform (or uniform). In most theoretical studies and applications, the distribution $F$ is assumed to be uniform \cite{HRP08,KRHP09,GRK05,Z15,SSH18,GMPS23}. However, it has recently been observed that a non-uniform random geometric graph may better fit real-world networks \cite{PBGKL23,G21}. Furthermore, RGGs with nonuniform $F$ exhibit properties that diverge significantly from those of RGGs with uniform $F$. Then a fundamental question is: whether the nodes of a graph generated from $\mathcal{G}_{n,m}(r_n,F)$ are uniformly distributed or not?  In this paper, we formulate this problem as a hypothesis testing problem and propose the first statistical test.

Let $U([0,1]^m)$ be the uniform distribution on the unit square $[0,1]^m$. Given $A\sim \mathcal{G}_{n,m}(r_n,F)$, we are interested in testing the following hypotheses
\begin{equation}\label{hyp1}
    H_0: F=U([0,1]^m),\hskip 2cm H_1: F\neq U([0,1]^m).
\end{equation}
Under the null hypothesis $H_0$, $F$ is the uniform distribution. Under the alternative hypothesis $H_1$, $F$ is a nonuniform distribution. This hypothesis testing problem is particularly challenging because, unlike the edges in classic random graphs, the edges in RGGs are not independent.  We will propose the first statistical test in the subsequent section.

\section{The proposed uniformity test}

A subgraph is a graph whose vertices and edges are subsets of those of a larger graph.
 Subgraphs play an important role in analyzing structures of networks. For example, \cite{BDER16} applied the number of triangles to detect geometry in random graph.
\cite{GL17,JKL18} used the number of short paths and cycles to test community structure in networks. Motivated by these applications, we plan to construct a test for (\ref{hyp1}) based on subgraphs. 

After careful analysis, we find that a function of short paths may serve as a powerful test. For $A\sim \mathcal{G}_{n,m}(r_n,F)$, define
\begin{equation}\label{gamman}
\Gamma_n=\mathbb{E}[A_{12}A_{23} A_{34}]\mathbb{E}[A_{12}]-\mathbb{E}[A_{12}A_{23}]^2.
\end{equation}
The first term of $\Gamma_n$ is the product of the expectation of 3-path and the expectation of edge. The second term is the square of the expectation of 2-path. If $A_{ij}$ are independent, it is straightforward to derive the order of $\Gamma_n$. However, the edges of $\mathcal{G}_{n,m}(r_n,F)$ are correlated. Determining the order of $\Gamma_n$ is a non-trivial problem, especially when $F$ is nonuniform. The following Proposition \ref{prop1} establishes the order of $\Gamma_n$. Under some mild conditions, $\Gamma_n=0$ under $H_0$ and $\Gamma_n\neq0$ under $H_1$.

\begin{Proposition} \label{prop1}
Let $A\sim \mathcal{G}_{n,m}(r_n,F)$. Suppose $r_n=o(1)$, $nr_n^m=\omega(1)$, and $F$ has continuous probability density function $f(x)$. If $F$ is uniform, then $\Gamma_n=0$. If $F$ is non-uniform, then
    \begin{eqnarray}\nonumber
     \Gamma_n=(2r_n)^{4m}\Delta_f+o(r_n^{4m}),     
    \end{eqnarray}  
   where $\Delta_f$ is a positive constant defined by
\begin{eqnarray}\label{deltaf4}
\Delta_f=\int_{[0,1]^m}f^{4}(x)dx\int_{[0,1]^m}f^{2}(x)dx-\left(\int_{[0,1]^m}f^{3}(x)dx\right)^2.
\end{eqnarray}
    
\end{Proposition}

From the Cauchy-Schwarz inequality, we deduce that
\begin{eqnarray}\nonumber
\left(\int_{[0,1]^m}f^{3}(x)dx\right)^2=\left(\int_{[0,1]^m}f^{1}(x)f^{2}(x)dx\right)^2\leq \int_{[0,1]^m}f^{2}(x)dx\int_{[0,1]^m}f^{4}(x)dx,
\end{eqnarray}
where equality holds if and only if 
\[f^{2}(x)=cf(x),\]
for some constant $c$. In this case,
\[f(x)=c.\]
Since $f(x)$ is the density function on $[0,1]^m$, then $f(x)=1$. Therefore, $\Gamma_n$ has order $r_n^{4m}$ if and only if $F$ is nonuniform. When $F$ is uniform, $\Gamma_n=0$. Then the quantity $\Gamma_n$ can distinguish $H_1$ from $H_0$.

Proposition \ref{prop1} leads to the consideration of the following pivotal test statistic.
\begin{eqnarray}\label{deftn}
T_{n}=\left(\sum_{i\neq j\neq k\neq l}A_{ij}A_{jk}A_{kl}\right)\left(\sum_{i\neq j}A_{ij}\right)-\left(\sum_{i\neq j\neq k}A_{ij}A_{jk}\right)^2.
\end{eqnarray}
The first term of $T_n$ is the product of the number of 3-paths and the number of edges. The second term is the square of the number of 2-paths.  
For large $n$, $\frac{T_{n}}{n^6}$ is expected to be close to $\Gamma_n$. Suitably scaled and centered, $T_{n}$ can be a powerful test statistic. 

In order to construct a test based on $T_n$, we need to derive its asymptotic distribution under the null hypothesis $H_0$. When $A_{ij}$ are independent, existing techniques can be used to derive the limiting distribution of $T_n$ (see \cite{GL17,JKL18,YLFS19} for examples). Nevertheless, the edges of $\mathcal{G}_{n,m}(r_n,F)$ are dependent.  Different techniques must be employed to determine the asymptotic distribution, which is specified in the following theorem. The proof is based on a novel application of the asymptotic theory of degenerate U-statistics with a kernel function that depends on $n$ \cite{FL96}.
 
\begin{Theorem}\label{mainthm}
Let $A\sim \mathcal{G}_{n,m}(r_n,F)$. Suppose $r_n=o(1)$ and $nr_n^m=\omega(1)$. Under the null hypothesis $H_0$, we have 
\begin{equation}\label{asymeq1}  
\frac{\sqrt{2}T_n}{n^5\bar{A}_n\hat{\sigma}_{n2}}\Rightarrow N(0,1),
\end{equation}
where 
\begin{equation}\label{hateqsi}
    \bar{A}_n=\frac{1}{n(n-1)}\sum_{i\neq j}A_{ij},\hskip 1cm
    \hat{\sigma}_{n2}^2=\frac{4\sum_{i_1\neq i_2\neq \dots \neq i_6}A_{i_1i_2}A_{i_2i_3}A_{i_3i_4}A_{i_4i_5}A_{i_5i_6}A_{i_6i_1}}{n^6}.
\end{equation}
\end{Theorem}

Note that the expected degree of a node in $A$ is proportional to $nr_n^{m}$ (see the proof of Proposition \ref{prop1}). If $r_n=o(1)$, the network is sparse, that is, the average degree of each node is of lower order than $n$. It is well-known that most real-world networks are sparse \cite{A17}. The proposed uniformity test is widely applicable to real-world network analysis. The condition $nr_n^m=\omega(1)$ implies that the average degree tends to infinity.  This condition is common in random geometric graph analysis \cite{GMPS23}.

The proof of Theorem \ref{mainthm} decomposes $T_n$ into a degenerate U-statistic and a small remainder.  In contrast to standard degenerate U-statistics, the kernel function in our paper's U-statistic depends on $n$. The standard asymptotic theory for degenerate U-statistics does not hold.  Instead, we utilize the asymptotic theory of degenerate U-statistic with kernel function dependent on $n$ in \cite{FL96}. According to Theorem \ref{mainthm}, the degenerate U-statistic in this case asymptotically follows the standard normal distribution. This result is different from the typical scenario where the kernel function is constant with respect to $n$, leading to an asymptotic distribution that is a sum of independent chi-square distributions.  Our method differs from prior approaches in network hypothesis testing problems \cite{L16,GL17,JKL18,YLFS19}.

 Based on Theorem \ref{mainthm}, we define the first test for hypotheses (\ref{hyp1}) as follows:
 \[\text{Reject $H_0$  if} \ \ \left|\frac{\sqrt{2}T_n}{n^5\bar{A}_n\hat{\sigma}_{n2}}\right|\geq Z_{\frac{\alpha}{2}},\]
where $Z_{\frac{\alpha}{2}}$ is the $(1-\frac{\alpha}{2})\%$ quantile of the standard normal distribution. We call this test the {\bf uniformity test}, which is summarized as Algorithm 1 in Table \ref{utt1}.

\begin{center}
\begin{table}[h] 
\begin{tabular}{  l } 
 \hline
\textit{Input:} Sparse network $A\sim \mathcal{G}_{n,m}(r_n,F)$.    \\ 
 \textit{Test}:  $H_0: F=U([0,1]^m)$, $H_1: F\neq U([0,1]^m)$.   \\ 
 1. Compute $T_n, \bar{A}_n, \hat{\sigma}_{n2}^2$ given in (\ref{deftn}) and (\ref{hateqsi}).\\
 2. Reject $H_0$  if $\left|\frac{\sqrt{2}T_n}{n^5\bar{A}_n\hat{\sigma}_{n2}}\right|\geq Z_{\frac{\alpha}{2}}$,\\
 \ \ \ \ where $Z_{\frac{\alpha}{2}}$ is the $(1-\frac{\alpha}{2})\%$ quantile of the standard normal distribution.  \\ 
 \hline
\end{tabular}\caption{Algorithm 1: The uniformity test.} \label{utt1}
\end{table}
\end{center}

According to Theorem \ref{mainthm}, the Type I error of the uniformity test converges to the nominal level $\alpha$ as $n$ goes to infinity.  In addition, the quantities $T_n$, $\hat{\sigma}_{n2}^2$ and $\bar{A}_n$ can be conveniently computed by using the adjacency matrix as follows:
\begin{equation}\label{tnmatrix0}
\bar{A}_n=\frac{\textbf{1}^TA\textbf{1}}{n(n-1)},
\end{equation}
\begin{equation}\label{tnmatrix}
T_n=\big[\textbf{1}^TA^3\textbf{1}-2(\textbf{1}^TA^2\textbf{1})+\textbf{1}^TA\textbf{1}-tr(A^3)\big]\big[\textbf{1}^TA\textbf{1}\big]-\big[\textbf{1}^TA^2\textbf{1}-tr(A^2)\big]^2,
\end{equation}
\begin{eqnarray}\nonumber
\hat{\sigma}_{n2}^2&=&\frac{4}{n^6}\Big[tr(A^6)-6tr(A^4)+5tr(A^3)-4tr(A^2)-3\textbf{1}^TA^3\textbf{1}+12\textbf{1}^TA^2\textbf{1}-3 \textbf{1}^T\big(\diag(A^3)\big)^2\\ \nonumber
&&+9 \textbf{1}^T\big(A^2\odot(A^2-J)\odot A\big)\textbf{1}-6\textbf{1}^T\big(A^2\odot(A^2-J)\odot (A^2-2J)\big)\textbf{1}\\ \label{hatmatrix} 
&&-2\textbf{1}^T\diag\big(A^2\odot(A^2-J)\odot (A^2-2J)\big)\Big].
\end{eqnarray}
Here, $M^T$ is the transpose of a matrix $M$. $tr(M)$ is the trace of $M$. $\textbf{1}^T=(1,1,\dots,1)$ is a row vector of length $n$. $J$ is a $n\times n$ matrix with all elements one. $\odot$ represents element-wise product of two matrices. $\diag(M)$ is a vector of the diagonal elements of the matrix $M$. It is easy to verify equation (\ref{tnmatrix0}) and equation (\ref{tnmatrix}). Equation (\ref{hatmatrix} ) can be derived from Theorem 4 in \cite{BR23}.

Next we analytically characterize the powers of the proposed tests.

\begin{Theorem}\label{mainpower}
Let $A\sim \mathcal{G}_{n,m}(r_n,F)$. Suppose $r_n=o(1)$ and $nr_n^m=\omega(1)$, and $F$ has continuous probability density function $f(x)$. Under the alternative hypothesis $H_1$, we have
\begin{eqnarray} \label{ppoereq1}
\frac{\sqrt{2}T_n}{n^5\bar{A}_n\hat{\sigma}_{n2}}=\Theta\left(\sqrt{n}\sqrt{nr_n^m}\Delta_f\right)(1+o_P(1)),
\end{eqnarray}
where $\Delta_f$ is given in (\ref{deltaf4}).
\end{Theorem}

Under the alternative hypothesis, $\Delta_f\neq0$. Since $nr_n^m=\omega(1)$ by assumption, then  $\sqrt{n}\sqrt{nr_n^m}\Delta_f=\omega(1)$. Therefore, the power of the proposed uniformity test tends to one as $n$ goes to infinity.

\subsection{An example}

As an example, we calculate $\Delta_f$ for the beta distribution with density function 
\begin{equation}\label{betadis}
   f_{a,b}(x)=\frac{x^{a-1}(1-x)^{b-1}}{B(a,b)}, \ \ x\in[0,1].
\end{equation}
Here $B(a,b)$ represents the beta function. The density function $f_{a,b}(x)$ is continuous on $[0,1]$ if $a\geq1$ and $b\geq1$. Especially, the uniform distribution corresponds to $a=1$ and $b=1$. It is easy to get that
\begin{eqnarray*}
\int_0^1f_{a,b}(x)^2dx&=&\frac{B(2a-1,2b-1)}{B(a,b)^2},\\
    \int_0^1f_{a,b}(x)^3dx&=&\frac{B(3a-2,3b-2)}{B(a,b)^3},\\ 
    \int_0^1f_{a,b}(x)^4dx&=&\frac{B(4a-3,4b-3)}{B(a,b)^4}.
\end{eqnarray*}
Let $F_{a,b}(x)$ be distribution function of the beta distribution with $a\geq1$ and $b\geq1$.  Define a $m$-dimensional distribution $F$ on $[0,1]^m$ as  \[F(x_1,x_2,\dots,x_m)=\prod_{t=1}^mF_{a,b}(x_t),\]
and \[f(x_1,x_2,\dots,x_m)=\prod_{t=1}^mf_{a,b}(x_t).\]
For $\mathcal{G}_{n,m}(r_n,F)$, it holds that
\begin{eqnarray*}\nonumber
\Delta_f&=&\int_{[0,1]^m}f^{4}(x)dx\int_{[0,1]^m}f^{2}(x)dx-\left(\int_{[0,1]^m}f^{3}(x)dx\right)^2\\
&=&\frac{\Big(B(2a-1,2b-1)\Big)^{m}\Big(B(4a-3,4b-3)\Big)^m-\Big(B(3a-2,3b-2)\Big)^{2m}}{B(a,b)^{6m}}.
\end{eqnarray*}
In general, $\Delta_f\neq0$. For instance,  $\Delta_f$ with $a=1$ has a simple expression as follows:
\begin{eqnarray*}\nonumber
\Delta_f&=&b^{6m}\frac{(3b-2)^{2m}-(2b-1)^m(4b-3)^m}{(2b-1)^m(4b-3)^m(3b-2)^{2m}}.
\end{eqnarray*}
Clearly, $\Delta_f\neq0$ if $b\neq1$.

\section{Simulations and real data applications}

In this section, we provide simulation studies and real data applications to support our theoretical results.

\subsection{Simulations}

We utilize simulation studies to assess the performance of the proposed uniformity test.
The Type I error rate is set at $\alpha=0.05$ throughout, and the networks are drawn from $\mathcal{G}_{n,m}(r_n,F)$. To assess the Type I error rate, we create 500 simulated networks based on the null hypothesis, using the $\mathcal{G}_{n,m}(r_n,F)$ model with uniform distribution $F$. Then we apply the proposed uniformity test to each network and note whether the null hypothesis is rejected.  The proportion of rejection among the 500 tests is the empirical Type I error. The empirical power calculation is identical, except that networks are sampled based on the alternative hypothesis.

In the first simulation, we consider $m=1$, $r_n\in\{0.030, 0.035, 0.040\}$, $n\in\{170,200,230\}$, and $F$ is the beta distribution with density given in (\ref{betadis}).  The results are summarized in Table \ref{simum1}. When $a=b=1$, the beta distribution is the uniform distribution. The column with $f_{1,1}$ contains the empirical Type I errors, and the powers are listed in the columns of $f_{a,b}$ with $a>1$ or $b>1$.  The empirical Type I errors are in line with the expected 0.05 level. The maximum power is close to one. An increase in $n$ or $r_n$ results in a corresponding increase in power.
When $a$ is fixed and $b$ increases, the power of the test becomes higher. The result remains valid when $b$ is kept constant and $a$ is increased. The proposed uniformity test performs very well.

\begin{table}[h]
\centering
  \begin{tabular}{|c|c| c| c|c|c|c|} 
 \hline
 & $n$ & $f_{1,1}$  & $f_{1,3}$ &$f_{1,5}$ &$f_{2,1}$ &$f_{3,1}$ \\  
 \hline
 \multirow{3}{5em}{$r_n=0.030$} 
 &  $170$ & 0.064   & 0.606&  0.860 & 0.300   &  0.596  \\  
 & $200$  & 0.052  &  0.810  & 0.962 & 0.496  & 0.824    \\  
 & $230$ &  0.046  &  0.924 &  0.990&  0.618  & 0.908 \\  
 \hline
 \multirow{3}{5em}{$r_n=0.035$} 
 &  $170$ & 0.054   & 0.664 & 0.898&   0.342  &   0.694  \\  
 & $200$  & 0.042   &  0.856  & 0.970& 0.522   & 0.850 \\  
 & $230$ &  0.032  & 0.940 & 0.996 & 0.656   & 0.944\\  
 \hline
 \multirow{3}{5em}{$r_n=0.040$} 
 &  $170$ &  0.046   & 0.706 & 0.910 & 0.360 &0.728\\  
 & $200$  & 0.040    & 0.890   & 0.976 & 0.570    &  0.888 \\ 
 & $230$ & 0.036 &  0.960  & 0.996  & 0.706   & 0.962\\  
 \hline
 \hline
 & $n$ & $f_{1,1}$  & $f_{2,5}$ &$f_{2,9}$ &$f_{4,3}$ &$f_{8,3}$ \\  
 \hline
 \multirow{3}{5em}{$r_n=0.030$} 
 &  $170$ & 0.064   &0.294&0.632& 0.254 &0.456\\  
 & $200$  & 0.052  & 0.526&0.804& 0.380& 0.618 \\  
 & $230$ &  0.046  &0.664&0.920 & 0.540& 0.806\\  
 \hline
 \multirow{3}{5em}{$r_n=0.035$} 
 &  $170$ & 0.054   &0.404&0.682&0.246 & 0.470 \\  
 & $200$  & 0.042  & 0.576&0.864& 0.442& 0.740\\  
 & $230$ &  0.032  &0.710&0.926 & 0.606&0.876\\  
 \hline
 \multirow{3}{5em}{$r_n=0.040$} 
 &  $170$ &  0.046   &0.424& 0.738&0.314&0.576\\  
 & $200$  & 0.040    & 0.642&0.880& 0.468 &0.744\\ 
 & $230$ & 0.036    &0.780&  0.970&0.676 & 0.900\\  
 \hline
 \end{tabular} 
\caption{Empirical Type I errors ($f_{1,1}$) and powers for $m=1$.}\label{simum1}
\end{table}

In the second simulation, we consider $m=2$, $r_n\in\{0.125, 0.130, 0.135\}$, $n\in\{170,200,230\}$ and $F$ is the distribution with density $f_{a_1,b_1}f_{a_2,b_2}$. The results are summarized in Table \ref{simum2}. The empirical Type I errors are shown in the column $f_{1,1}f_{1,1}$, and the powers are reported in the columns with $f_{a_1,b_1}f_{a_2,b_2}$ for $a_1>1$ or $b_1>1$ or $a_2>1$ or $b_2>1$. The Type I errors closely match the expected rate of 0.05. The power for non-uniform two components is notably higher than the power with one  non-uniform component and one uniform component. The maximum power is almost one. The power increases as $n$ or $r_n$ increases.  The proposed uniformity test shows excellent performance.

\begin{table}[h]
\centering
  \begin{tabular}{|c|c| c| c|c|c|c|} 
 \hline
 & $n$ & $f_{1,1}f_{1,1}$  & $f_{1,1}f_{1,2}$
   & $f_{1,2}f_{1,2}$ & $f_{1,2}f_{2,1}$ & $f_{2,2}f_{2,2}$ \\  
 \hline
 \multirow{3}{5em}{$r_n=0.120$} 
 &  $170$ &0.050& 0.210& 0.830& 0.816&0.538\\   
 & $200$  &0.058&0.372&0.940&0.940& 0.754\\   
 & $230$ & 0.046&0.508&0.976& 0.974 & 0.870\\   
 \hline
 \multirow{3}{5em}{$r_n=0.125$} 
 &  $170$ & 0.052&0.196 &0.854& 0.834& 0.554 \\  
 & $200$  &0.048& 0.352& 0.936& 0.944& 0.756\\  
 & $230$ & 0.046&0.506&0.990& 0.980& 0.886\\  
 \hline
 \multirow{3}{5em}{$r_n=0.130$} 
 &  $170$ &0.040 &0.222&0.848&0.844 & 0.664 \\   
 & $200$  &0.038&0.360&0.956& 0.956& 0.786\\   
 & $230$ &0.032&0.566&0.998&0.986& 0.918\\    
 \hline
 \hline
 & $n$ & $f_{1,1}f_{1,1}$  & $f_{1,1}f_{2,3}$  &$f_{2,3}f_{2,3}$ &$f_{2,3}f_{3,2}$ &$f_{3,3}f_{3,3}$ \\  
 \hline
 \multirow{3}{5em}{$r_n=0.120$} 
 &  $170$ & 0.050& 0.144&0.774& 0.770 &  0.816\\      
 & $200$  & 0.058 &0.230&0.912& 0.914 & 0.956 \\         
 & $230$ &0.046& 0.330&0.976&0.982& 0.982\\   
 \hline
 \multirow{3}{5em}{$r_n=0.125$} 
 &  $170$ &0.052 &0.142&0.812&0.806&0.842\\   
 & $200$  & 0.048 &0.222&0.922& 0.926& 0.958\\    
 & $230$ &0.046&0.366&  0.976& 0.988&0.984\\   
 \hline
 \multirow{3}{5em}{$r_n=0.130$} 
 &  $170$ & 0.040&0.150&0.848& 0.838& 0.846\\    
 & $200$  & 0.038&0.238&0.928& 0.936& 0.960\\   
 & $230$ & 0.032& 0.372& 0.988& 0.990& 0.994\\   
 \hline
 \end{tabular} 
\caption{Empirical Type I errors ($f_{1,1}f_{1,1}$) and powers  for $m=2$.}\label{simum2}
\end{table}

\subsection{Real data applications}

The proposed uniformity test was implemented on multiple real-world network datasets in \cite{HDATA}. These networks comprise three types of networks: social network, brain network and animal social network. The social network represents social interactions between entities. The brain network models fiber tracts that connect one brain region node to another. Table \ref{realdata1} provides information on the networks, including the number of nodes and edge densities. The first network listed is a social network, with the second being the brain network and the last four being animal social networks.

We test the null hypothesis that the distribution of nodes in each network is uniformly distributed. We perform the proposed uniformity
test and calculate its p-value. The results are shown in Table \ref{realdata1}. The highlighted are p-values less than Type I error $\alpha=0.05$. At significance level $\alpha=0.05$, we reject the null hypothesis for the three networks `mammalia-dolphin-floridatravel', `mammalia-dolphin-florida-overall' and `reptilia-tortoise-network-fi'. The p-value for the first three networks is greater than the significance level, so we fail to reject the null hypothesis. These results indicate that nodes of some networks are distributed uniformly, and others exhibit non-uniform distributions. Before modeling a network using a RGG, it is essential to test whether nodes are uniformly distributed across the metric space.

\begin{table}[h]
\centering
  \begin{tabular}{|c| c| c|c|} 
 \hline
Network & $n$ & density & p-value  \\  
 \hline
firm-hi-tech & 33 & 0.235 & 0.135   \\  
macaque-rhesus-brain-1  &242 & 0.140 &   0.289\\ 
aves-weaver-social  & 445  & 0.013  &   0.774 \\
 mammalia-dolphin-floridatravel  & 188   & 0.058  & {\bf 0.001}\\
 mammalia-dolphin-florida-overall &291& 0.075 &  {\bf 0.000}  \\
 reptilia-tortoise-network-fi &787& 0.004 & {\bf 0.000}\\
 \hline
 \end{tabular} 
\caption{P-values of the proposed uniformity test. }\label{realdata1}
\end{table}


\begin{thebibliography}{9}

\bibitem{A17} Abbe, E. (2017).
Community detection and stochastic block models: recent developments. 
\textit{Journal of Machine Learning Research}, \textbf{18}, 1-86.




\bibitem{ARS01}
Agiwal, M., Roy, A., Saxena, N. (2016). Next generation 5G wireless networks: A comprehensive survey. \textit{IEEE communications surveys and tutorials}, 18(3), 1617-1655.


\bibitem{BR23}
Barika,S. and Redd, S. (2023).
Number of cycles of small length in a graph,
\textit{AKCE International Journal of Graphs and Combinatorics}, 20,2, 134–147.




\bibitem{BKL19}
Bringmann,K., Keusch, R. and Lengler, J. (2019).
Geometric inhomogeneous random graphs, \textit{Theoretical Computer Science}, 760, 35-54.





\bibitem{BB24}
Bangachev,K. and Bresler, G.(2024).
Detection of $L_{\infty}$ geometry in random geometric graphs:suboptimality of triangles and cluster expansion. \textit{Proceedings of Machine Learning Research}, 247:1–71.





\bibitem{BDER16}
Bubeck, S., Ding, J., Eldan, R., and Racz, M.(2016)
Testing for high‐dimensional geometry in random graphs. \textit{Random Structures \& Algorithms}, 49,503-532.



    
    \bibitem{BC23}Badiu, M.-A. and Coon, J. P.(2023),
    Structural complexity of one-dimensional random geometric graphs, \textit{IEEE Transactions on Information Theory},  69(2),794-812.
    
    
 
    
    \bibitem{CC12} Caldarelli, G.and Catanzaro, M.(2012),
    \textit{Networks: A Very Short Introduction},Oxford University Press.






\bibitem{CGL16}
 Chiasserini, C.F.,  Garetto, M. and  Leonardi, E. (2016). Social network de-anonymization under
scale-free user relations. \textit{IEEE/ACM Transactions on Networking} 24 (6):3756–3769.

\bibitem{CBL16}
Charitou, T., Bryan, K. and Lynn, D.(2016).
Using biological networks to integrate, visualize and analyze genomics data,\textit{Genetics Selection Evolution}, 48,27.




\bibitem{CLX01}
Cao, L., Li, L., Huang, Z., Xia, F., Huang, R., Ma, Y., Ren, Z. (2023). Functional network segregation is associated with higher functional connectivity in endurance runners. Neuroscience Letters, 812, 137401


\bibitem{CV1}
Costanzo, M., VanderSluis, B., Koch, E. N., Baryshnikova, A., Pons, C., Tan, G., Boone, C. (2016). A global genetic interaction network maps a wiring diagram of cellular function. Science, 353(6306), aaf1420.







\bibitem{DC23}
Duchemin, Q., De Castro, Y. (2023). Random geometric graph: some recent developments and perspectives. \textit{High Dimensional Probability IX. Progress in Probability} Birkhäuser, Cham.  




\bibitem{HRP08}
Desmond J. Higham, Marija Rašajski, Nataša Pržulj (2008). Fitting a geometric graph to a protein–protein interaction network, \textit{Bioinformatics}, 24(8), 1093–1099.









\bibitem{FL96}
Fan, Y. and Li, Q. (1996). Consistent model specification tests: omitted variables and semiparametric functional
forms.\textit{Econometrica},  64, 4,865-890.




 

\bibitem{GMPS23}
Galhotra, S., Mazumdar, A. , Pal, S.,  Saha, B.(2023).
Community recovery in the geometric block model, \textit{Journal of Machine Learning Research} 24: 1-53.

\bibitem{GRK05}
Goel, A., Rai, S., Krishnamachari, B.(2005). Monotone properties of random geometric graphs have sharp thresholds. \textit{Ann. Appl. Probab.} 15 (4) 2535-2552.

\bibitem{G21}
Ganesan, G., Robust paths in random geometric graphs with applications to mobile networks, \textit{2021 International Conference on COMmunication Systems \& NETworkS (COMSNETS)}, Bangalore, India, 2021, pp. 119-123.
    

\bibitem{GL17} Gao, C. and Lafferty, J. (2017).
Testing for global network structure using small subgraph statistics.
\url{https://arxiv.org/pdf/1710.00862.pdf}    





\bibitem{HM09}
Han, G. and Makowski, A.(2009). One-dimensional geometric random graphs with
nonvanishing densities—Part I: A strong zero-one
law for connectivity. \textit{IEEE Transactions on information theory}, 55(12),5832-5839.

\bibitem{HM12}
Han, G. and Makowski, A.(2012). One-dimensional geometric random graphs with
nonvanishing densities—Part II: a very strong zero-one
law for connectivity. \textit{Queueing Syst}, 72, 103-138.




\bibitem{JKL18} 
Jin, J., Ke, Z., Luo, S.(2018).
Network global testing by counting graphlets,  \textit{PMLR},
2333-2341.




\bibitem{KRHP09}
Kuchaiev O, Rašajski M, Higham DJ, Pržulj N (2009). Geometric de-noising of protein-protein interaction networks. \textit{PLOS Computational Biology} 5(8): e1000454.




















  

\bibitem{LLR15}
Liu, L., Lei, J. and Roeder, K. (2015).
Network assisted analysis to reveal the genetic basis of autism,  \textit{Ann Appl Stat.} 2;9(3):1571–1600.




\bibitem{L16} Lei, J. (2016).
A goodness-of-fit test for stochastic block models.
\textit{Annals of Statistics}, \textbf{44}, 401-424. 



\bibitem{BJ23}
Mihai-Alin Badiu, Justin P. Coon(2023).
Structural complexity of one-dimensional random
geometric graphs. \textit{IEEE Transactions on information theory}, 69(2),794-812.




\bibitem{MMFE22}
 Martinez-Martinez, C. T., Mendez-Bermudez, J. A., Rodrigues, Francisco A., Estrada, Ernesto (2022).
Nonuniform random graphs on the plane: A scaling study,\textit{ Physical Review E},105, 034304

\bibitem{N03}
Newman, M. (2003). The Structure and Function of Complex Networks. \textit{SIAM review} 45, (2), 167–256.




\bibitem{HDATA} Network Repository: https://networkrepository.com/




\bibitem{BM08}
 O'Malley, A. J. and Marsden,P. V. (2008). The analysis of social networks, \textit{Health Serv Outcomes Res Methodol}, {\bf 8}, 222-269.







\bibitem{PBGKL23}
Paolino, R., Bojchevski, A., Gunnemann, S., Kutyniok, G. and Levie, R.(2023).
Unveiling the sampling density
in non-uniform geometric graphs, \textit{ICLR 2023}




\bibitem{RPF13}
Rinaldo, A., Petrovic, S. and Fienberg, S.(2013).
 Maximum Likelihood Estimation in the $\beta$-Model. \textit{Annals of Statistics}, 41(3), 1085-1110.
    


\bibitem{SBL13}
Simpson, S., Bowman F. and Laurienti, P.(2013). Analyzing complex functional brain networks: Fusing statistics and network science to understand the brain,\textit{Statistics Surveys}, 7: 1-36.



\bibitem{SSH18}
Solovey, K., Salzman, O. and Halperin, D.(2018).
New perspective on sampling-based motion planning via random geometric graphs.\textit{The International Journal of Robotics Research},37(10):1117-1133.




\bibitem{SHL11}
Simpson SL, Hayasaka S, Laurienti PJ. Exponential random graph modeling for complex brain networks. \textit{PLoS One}, 6(5):e20039.

   
\bibitem{YLFS19} Yuan, M., Liu, R., Feng, Y. and Shang, Z.(2022). Testing community structures for hypergraphs.
\textit{Annals of Statistics}, 50(1): 147-169.

    

\bibitem{Z15}
Zhao, J. (2015). The absence of isolated node in geometric random graphs, \textit{53rd Annual Allerton Conference on Communication, Control, and Computing (Allerton)}, Monticello, IL, USA, 2015, pp. 881-886.


    
 


\bibitem{ZLKG14}
Zhang, W., Lim, C., Korniss, G. et al. Opinion dynamics and influencing on random geometric graphs. \textit{Sci Rep} 4, 5568 (2014).



 


\end{thebibliography}
\end{document}